\documentclass{epl}
\usepackage{graphicx}
\usepackage{dcolumn}
\usepackage{amsmath}
\usepackage{amssymb}
\usepackage{bm}
\usepackage{epsfig}
\usepackage{psfrag}

\def\R{{\bf R}}
\def\Itens{\mbox{\sffamily\bfseries I}}

\def\nabbold{\mbox{\boldmath $\nabla$\unboldmath}}

\def\beq{\begin{equation}}
\def\eeq{\end{equation}}
\def\bea{\begin{eqnarray}}
\def\eea{\end{eqnarray}}
\title{Flow-induced currents in nanotubes: a Brownian dynamics approach}
\shorttitle{Flow-induced currents in nanotubes: Brownian dynamics}
\author{Moumita DAS\inst{1,2}, Sriram RAMASWAMY\inst{1}, A. K. SOOD\inst{1}
\and G. ANANTHAKRISHNA\inst{3}}
\shortauthor{M. Das \etal}
\institute{
  \inst{1} Department of Physics, Indian Institute of Science, Bangalore 560012, India\\
  \inst{2} Division of Engineering and Applied Sciences, Harvard University, Cambridge, MA-02138,USA.\\
  \inst{3} Materials Research Centre, IISc, Bangalore 560012, India.
}
\begin{document}
                                                                                
\maketitle

\begin{abstract}
Motivated by recent experiments \cite{shankar} reporting that carbon nanotubes 
immersed in a flowing fluid displayed an electric current and voltage, 
we numerically study the behaviour of a collection of Brownian particles in a 
channel, in the presence of a flow field applied on similar but slower 
particles in a wide chamber in contact with the channel.
For a suitable range of shear rates, we find that the flow field induces a
unidirectional drift in the confined particles, and is stronger for narrower
channels. The average drift velocity initially rises with increasing shear
rate, then shows saturation for a while, thereafter starts decreasing,
in qualitative agreement with recent theoretical studies \cite{shankar2}
based on Brownian drag and ``loss of grip''.
Interestingly, if the sign of the interspecies interaction is reversed,
the direction of the induced drift remains the same, but the
flow-rate at which loss of grip occurs is lower, and the
level of fluctuations is higher.
\end{abstract}


Recent experiments show that liquid flow past single walled carbon nanotube 
(SWNT) bundles generates voltage \cite{shankar} and 
electric current \cite{shankar2} in the nanotube, along the direction of flow. 
The current and voltage are found \cite{shankar,shankar2} to be highly 
sublinear functions of the fluid flow rate. 
The direction of flow-induced current relative to fluid flow is determined 
\cite{shankar,shankar2} by the nature of the ions in the vicinity of the nanotubes. 
The experiments in \cite{shankar, shankar2} also showed that the one-dimensional nature of the SWNTs was essential for the 
generation of a net electrical signal in the sample: experiments on graphite 
did not generate a measurable signal and 
those on multiwalled carbon nanotubes produced a signal approximately 10 times 
weaker than that for the single walled 
nanotubes, the dimensions of the sensing element and flow speeds of the ionic 
liquid remaining same. Electrokinetic \cite{cohen} and phonon-wind \cite{kral} based
explanations give a linear current versus flow rate dependence. There is also a recent proposal \cite{tosatti}
involving stick-slip and barrier hopping of ions giving a sublinear dependence.
Ghosh {\it et al.} \cite{shankar2} explain the phenomenon as follows: 
Thermal fluctuations in the ionic charge density in the fluid near the nanotube produce 
a stochastic Coulomb field on the carriers in the nanotube. At thermal equilibrium, 
i.e., if there is no mean fluid flow, the fluctuation-dissipation theorem 
tells us that the power spectral density 
of these fluctuations contributes to the frictional drag on the carriers. 
The total friction or resistivity of the carriers is thus the sum 
of contributions from the ambient ions and from phonons or other mechanisms 
intrinsic to the nanotube. 
An imposed fluid flow advects the ions. The carriers in the nanotube then have to 
move at a speed determined by balancing the drag forces due to the ions and the 
nanotube. However, increasing the fluid velocity also 
Doppler-shifts the autocorrelation 
function of the ionic charge density, hence speeding up its time-decay 
as seen by the carriers and reducing the drag due to the ions. This results in 
a saturation of the flow-induced voltage and current.   
The calculation of \cite{shankar2} was done in a simplified framework, where the 
ions were assumed to move with a single mean velocity. In the experiment 
the velocity of fluid (and ions) increases with distance from the nanotube surface. 
Ions nearest the nanotube contribute the strongest Coulomb field but move the slowest. 
It would be interesting, first, to see if the form of the dependence of current 
on flow-speed remains intact when these competing tendencies are taken into account.  
Secondly, it remains to be seen if the theory of \cite{shankar2} produces 
a weaker effect when extended to flow past a higher dimensional conductor. 
Lastly, only the two-point correlations of the ionic Coulomb field enter the 
(Gaussian) treatment of \cite{shankar2}, which means that changing the sign of 
the ionic charge can have no effect. Testing the importance of non-Gaussian 
fluctuations is important, but hard to do analytically. All three questions 
are easily resolved in a numerical simulation.   

Motivated by these considerations, we carry out a Brownian dynamics study 
on a model system (Figs. \ref{modelfigure}) 
consisting of two types of particles, (i) particles `A'
of low diffusivity in a chamber (ions) and (ii) particles `B' of comparatively 
high diffusivity 
(carriers) confined in an adjoining channel (the nanotube) of depth $W_B$ 
much smaller than that of the chamber.
Both `A' and `B' particles are assumed to be charged particles with pairwise 
screened Coulomb interactions.  
Our simulations 
implement the theoretical ideas
of \cite{shankar2} with an important difference: the ions flowing past the 
nanotube have a strong velocity gradient in the vicinity of the nanotube, as 
shown in Fig. \ref{modelfigure}.
For a suitable range of shear rates (magnitude of velocity gradient), 
we find that an imposed flow of the `A' particles indeed induces a
unidirectional drift in the `B' particles in the channel. 
The average drift velocity initially rises with increasing shear
rate, then shows saturation for a while, and thereafter starts decreasing,
in qualitative agreement with the theoretical arguments of \cite{shankar2}.
We further observe that the induced drift decreases substantially when $W_B$ is increased,
underlining as in the experiments of \cite{shankar}, the central role of reduced dimensionality. 
Interestingly, if the sign of the A-B interaction is reversed,
the direction of the induced drift remains the same, in agreement 
with \cite{shankar2}, but the
flow-rate at which crossover to saturation occurs is lower, and the
level of fluctuations is higher. 

\section{Model and Results}
\begin{figure}
\onefigure[width=8cm]{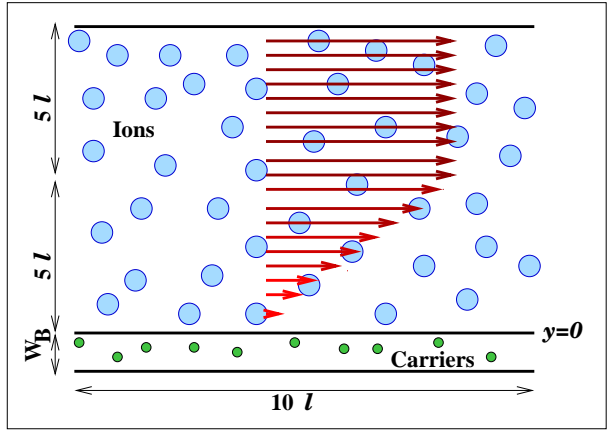}
\caption{\label{modelfigure} A schematic diagram of the model.}
\end{figure}

Let us now present our study and its results in more detail. We consider for simplicity a two dimensional ($x$-$y$) 
system consisting of two species of particles, say, A representing the ions in the chamber 
(of dimension $10 \ell \times 10 \ell$, ) and B, representing the carriers in the 
narrow channel 
(of dimension $10 \ell \times \epsilon \ell$, $\epsilon \leq $2), 
with $\ell = (\rho_A)^{-1/2}$, 
where $\rho_A$ is the mean number density of ions. We work with periodic boundary conditions in
the $x$ direction, and hard walls at $y=10 \ell$, $y=0$ and $y=- \epsilon \ell$. The flow field imposed
on A particles is plane Couette, with velocity in the $x$ direction and gradient in the $y$ direction 
till a certain separation  between the walls ($5 \ell$), beyond which the velocity is uniform (Fig. \ref{modelfigure}).
This flow geometry effectively models the flow field near a small obstacle in bulk flow, such that gradients are 
concentrated in the boundary layer on the surface of the obstacle.

There is no flow field imposed on B particles. Both A and B particles are overdamped and the pairwise
interactions are screened Coulomb in nature. Further B particles are 100 times more diffusive than their A
counterparts. 
The positions $\R(t)$ ($={x(t) , y(t)}$),  
evolve according to overdamped Langevin equations, with independent Gaussian, zero-mean,
thermal white noise sources ${\bf h_A}$ (or ${\bf h_B}$), interparticle forces $\nabbold V$ 
from the pair potentials, 
and for the ions an additional force due to the flow field $\dot \gamma \hat{\bf x}$.
Let us nondimensionalise our variables as follows:
scale all lengths by $\ell$, energy by Boltzmann's constant $k_B$ times
temperature $T$ (and hence force by $k_BT/\ell$),
and time by the time $\tau$ taken by a carrier to traverse a distance $\ell$.
Then our nondimensional discretised Langevin equations are for a given type ($A$ or $B$) of particles,
\begin{equation}
\label{e.langnondim}
{\R_A}(t + \delta t) = {\R_A}(t) + D_A [ \dot \gamma \hat{\bf x} - \nabbold V_{AA} -\nabbold V_{AB}] \delta t + \sqrt{2 D_A \delta t} \,\, {\bf h_A}(t), 
\end{equation}
\begin{equation}
{\R_B}(t + \delta t) = {\R_B}(t) - D_B \nabbold V_{BA} \delta t   + \sqrt{2 D_A \delta t} \,\, {\bf h_B}(t)
\end{equation}
where $D_A(=1)$ and $D_B(=100)$ are the nondimensionalised Brownian diffusivities for the ions and carriers respectively and  obey the fluctuation dissipation relation, i.e., for example for $A$ particles, $\langle {\bf h_A}^i(0) {\bf h_A}^j(t) \rangle = 2 \Itens \delta^{ij} \delta(t)$, where $\Itens$ is the unit tensor 
and $i,j$ label particle indices.
The pair potentials have the screened Coulomb form
$V(r) = (U/r) \exp(-\kappa r)$ (where $U$ is a coupling coefficient of the DLVO form \footnote{This DLVO coefficient is given by $Z_i Z_j e^2 exp(\kappa (\sigma_i + \sigma_j)) /(\epsilon (1+\kappa \sigma_i)(1+\kappa \sigma_j))$, where $\epsilon$ is the dielectric constant of the solvent, $Z_i e$ is the charge on and $\sigma_i$ is the radius of the $i^{th}$ particle. In our simulations, the dimensionless value of the $\kappa$ independent part of $U$ 
is $\approx 15.3$,  $\sigma_{B}/\sigma_{A}= 0.1$ and $2 \sigma_A/\ell \approx 0.455$.}.)
at interparticle separation ${\bf r}$, with different $\kappa$'s for $AA$ and $AB$ interaction, while there is no $BB$ interaction.
Also both the $A$ and $B$ particles carry charges of identical sign and magnitude.
The dimensions of the box containing $A$ particles is $L = 10 $ 
and $W = 10 $,
whereas for B particles $L=10$, and $W=\epsilon (\epsilon \le 2)$.
Keeping $W_A$=10 and $\kappa_{AB} \ell$=2 fixed, we monitor the behavior of the system by changing:
(1) Width of the channel containing the carriers $W_B$, (2) Ion-ion interaction strength, $\kappa_{AA} \ell$ and (3) sign of the $A-B$ i.e. ion-carrier
interaction. We also explore the effect of changing the screening length of 
the ion-carrier ($A-B$) interaction, keeping other parameters constant.
We have studied the behavior of this model for a system with $N_A= 100$ particles and $\rho_A/\rho_B=1$ and $\rho_A/\rho_B=1/2$, $\rho_A$ and $\rho_B$ being the number densities for the ions and carriers respectively. The results presented here
pertain to the latter case, the observed behavior being qualitatively same for 
both cases.

\begin{figure}
\onefigure[width=8cm]{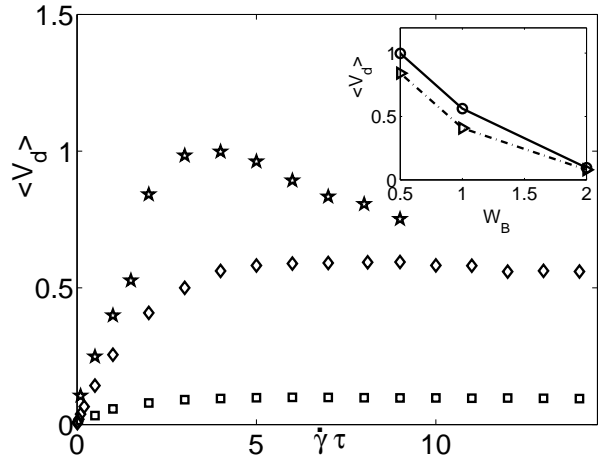}
\caption{\label{diacAPR3kappaa1d1a} The induced drift of the carriers as a function of the ionic flow speed
for different $W_B$ (stars, diamonds and squares  correspond to $W_B=0.5$,$W_B=1.0$ and $W_B=2.0$ respectively) for $\kappa_{AA} \ell$=1. The inset shows the
average drift velocity $\langle v_d \rangle$ as a function of $W_B$, for $\dot \gamma \tau$ = 4.0 (circles) and 2.0 (triangles) respectively.}
\end{figure}

For $W_B= 2$, we observe that the carriers
do have a unidirectional drift in the direction of the flow. 
The average drift velocity of the carriers initially increases with the flow rate, 
reaches a maximum and then decreases at large values of the flow rate as seen in Fig.
\ref{diacAPR3kappaa1d1a}. This is broadly in agreement with the ideas of \cite{shankar2}.
On further decreasing $W_B$ to first $1.0$ and then $0.5$ (Fig.\ref{diacAPR3kappaa1d1a}), 
we find that the drift velocity induced is progressively larger than for $W_B = 2.0$, 
and the saturation occurs at a lower value of the flow rate.

\begin{figure}
\twofigures[width=7cm]{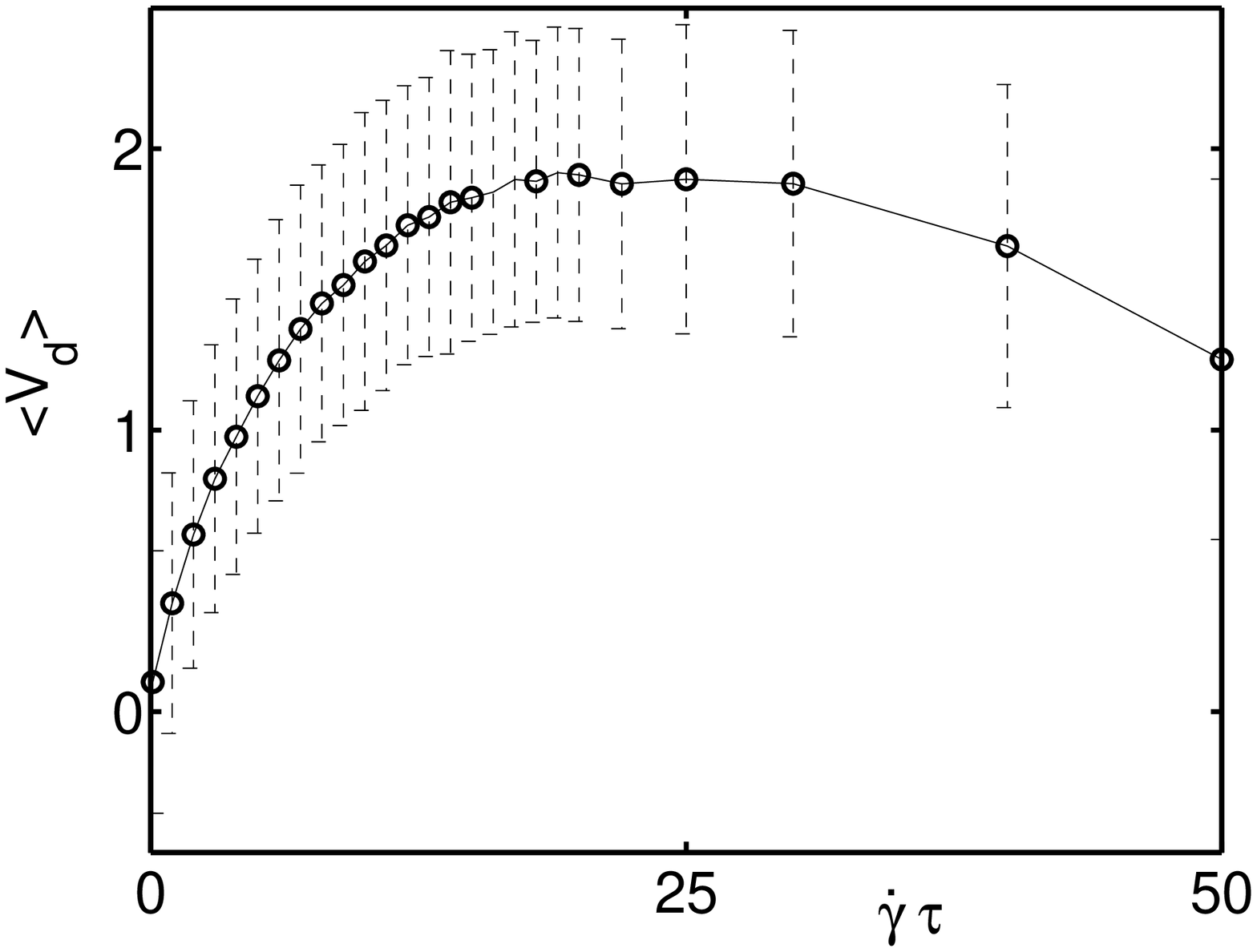}{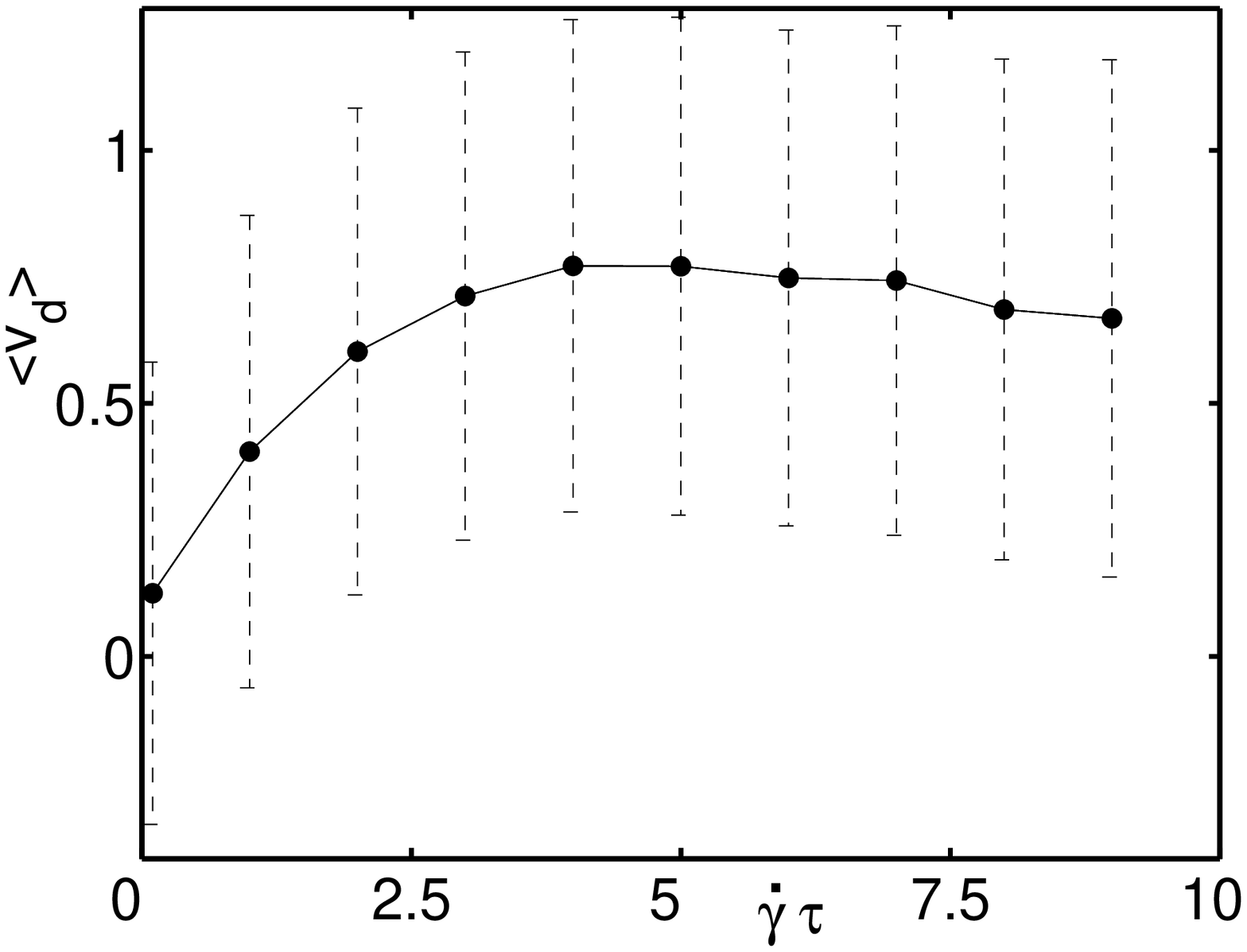}
\caption{\label{repulwbpt5kappaapt2} Induced drift as function of flow speed
for $\kappa_{AA} \ell$=0.2, $W_B$=0.5. The vertical dashed lines show the standard deviation from the mean drift velocity. }
\caption{\label{repulwbpt5kappaa4} Induced drift as function of flow speed for $\kappa_{AA} \ell$=4,  $W_B$=0.5. Note that the velocity weakening takes place at a much smaller shear rate
than in Fig.\ref{repulwbpt5kappaapt2}. The vertical dashed lines show the standard deviation from the mean drift velocity.}
\end{figure}

Now keeping the  wall to wall distance fixed at $W_B=0.5 $, 
we study the behavior of the system at
very small ($\kappa_{AA} \ell$=0.2) (Fig. \ref{repulwbpt5kappaapt2})and 
large ($\kappa_{AA} \ell$=4) ion-ion interaction screening strength 
(Fig. \ref{repulwbpt5kappaa4}).
A smaller value of $\kappa \ell$ corresponds to stronger ion-ion interactions and it is observed
that in such a case the induced effect on the carriers is stronger and is sustained till a larger value 
of the ionic flow speed at which the weakening sets in.

\begin{figure}
\twofigures[width=7cm]{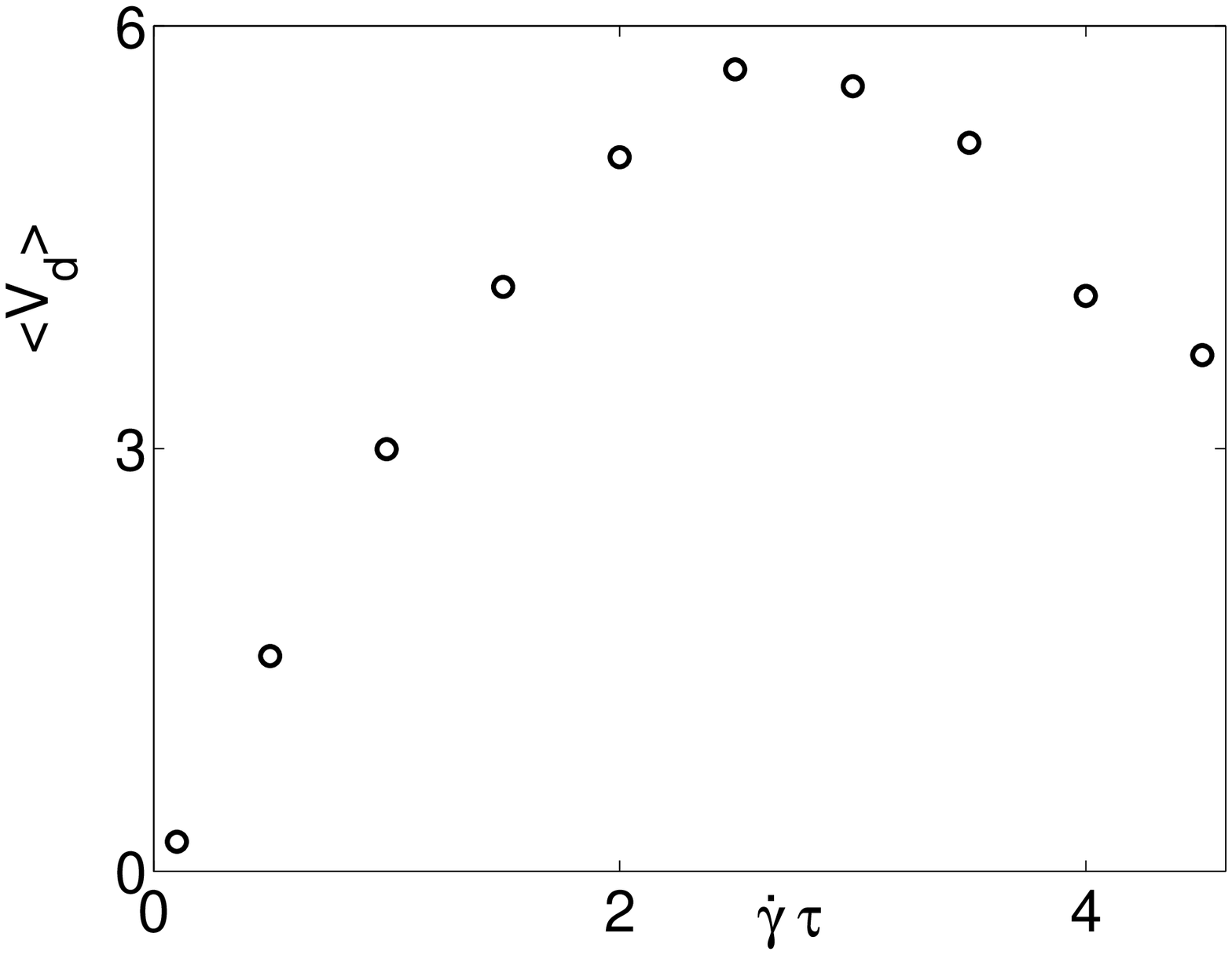}{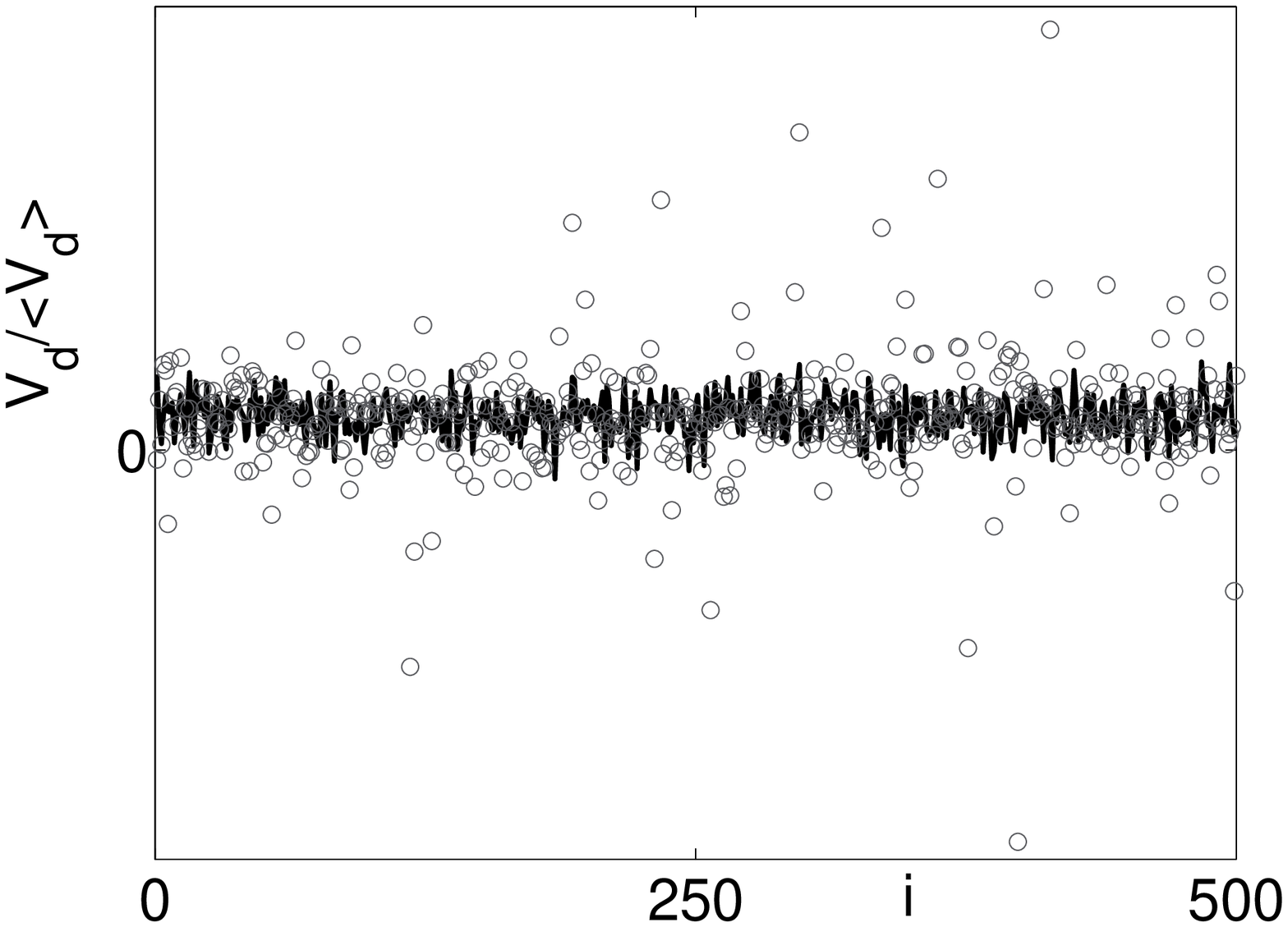}
\caption{\label{attractAB} Induced drift as a function of the ionic flow speed
for $\kappa_{AA} \ell$=1, $W_B$=1, with attractive ion-carrier interaction.}
\caption{\label{attractiverepulsivefluc} Drift velocities $V_d$ of the carriers at different times (represented by the index 'i') scaled by the average value $\langle V_d \rangle$ at $\kappa_{AA} \ell$=1, $W_B$=1, for repulsive (lines) and attractive (circles) ion-carrier interactions.}
\end{figure}

We now look at the scenario where the two species A (ions) and B (carriers) have opposite charges (Fig. \ref{attractAB}).
The wall to wall distance is kept fixed at $W_B=1$. We observe that the induced drift is stronger than
when both A and B particles carry like charges, at the same $W_B$. The fluctuations in 
the drift velocity are however much larger than for the case of like charges (Fig. \ref{attractiverepulsivefluc}): 
the ratio of standard deviation to mean is $\sim 4$ times that for the repulsive case. 
This can be explained as follows: 
The attraction of the carriers for the ions brings both close to the wall separating them, from time to time, enhancing 
the flow induced current. Subsequently the presence of the walls pushes them apart, reducing the induced current again. 
This implies large fluctuations in the current. For like charged ions and carriers both the Coulomb repulsion and the 
walls act in tandem to keep them apart, thus the intermittent enhancement and reduction in current does not occur.

\begin{figure}\onefigure[width=8cm]{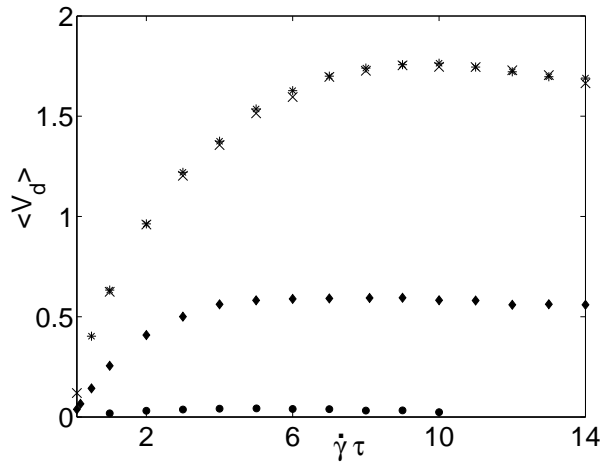}
\caption{\label{diffkappaABinverse}The flow induced drift velocity of B particles as a function of the flow speed of A particles, for $\kappa_{AB}^{-1}= 0.1 \ell$ (circles), $0.5 \ell$ (diamonds), $5 \ell$ (cross) and $10 \ell$ (plus), for fixed $W_B$=1.0 and $\kappa_{AA} \ell=1$.}
\end{figure}

In order to understand the dependence of the effect on the strength of A-B interaction, we monitored the system for different values of the screening length
corresonding to the A-B interaction (Fig. \ref{diffkappaABinverse}) keeping the channel width $W_B$ and A-A interaction strength constant. We find
that the effect increases on increasing the screening length $\kappa_{AB}^{-1}$
and is maximum 
where $\kappa_{AB}^{-1}$ is comparable to the width of the channel $W_B$; therafter it saturates and doesn't show a marked increase.
This shows that the contribution to the flow induced effect indeed chiefly comes from the A particles (or ions) present in a boundary layer of order of the width of the channel at the interface of the chamber and the channel.

\begin{figure}
\twofigures[width=7cm]{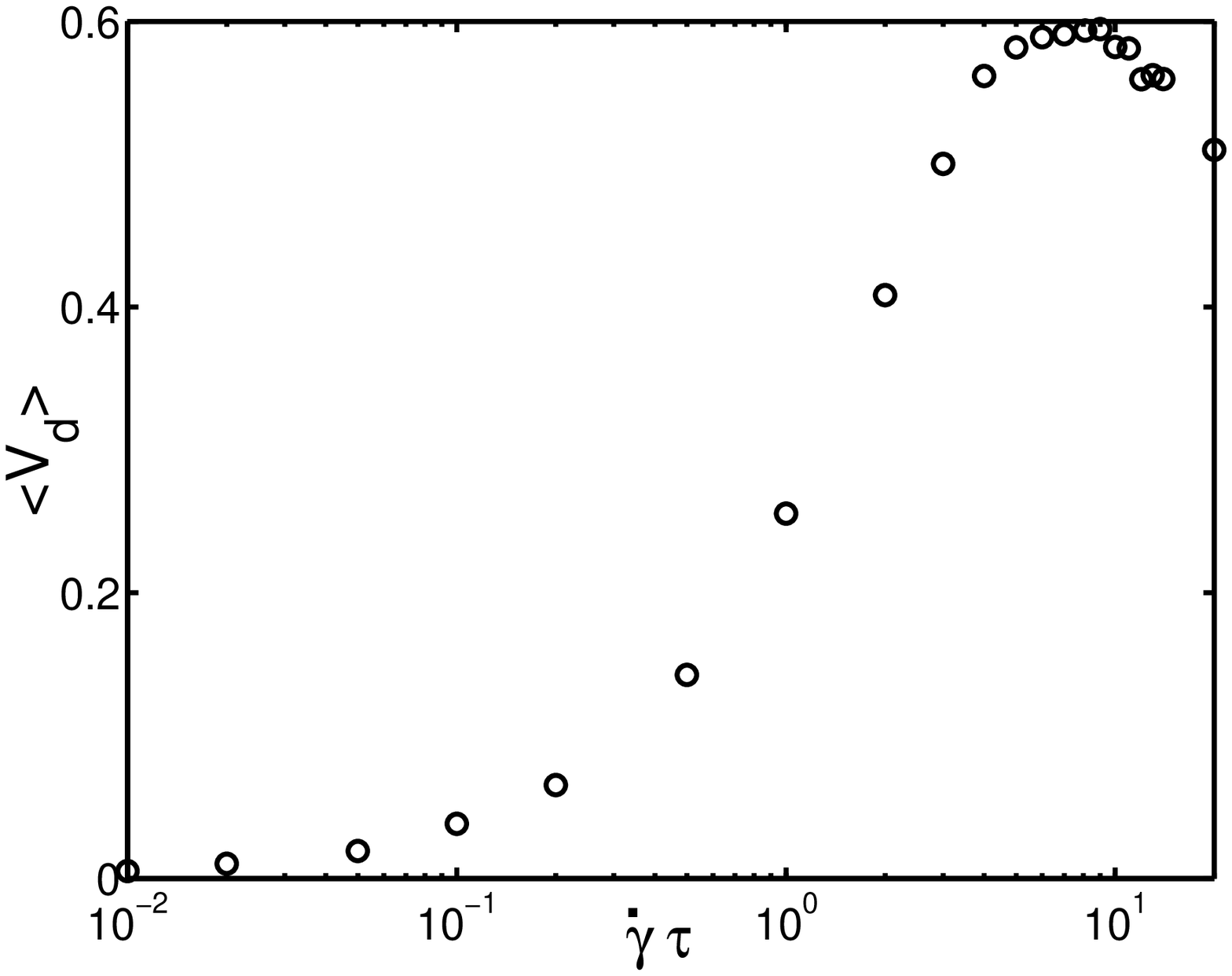}{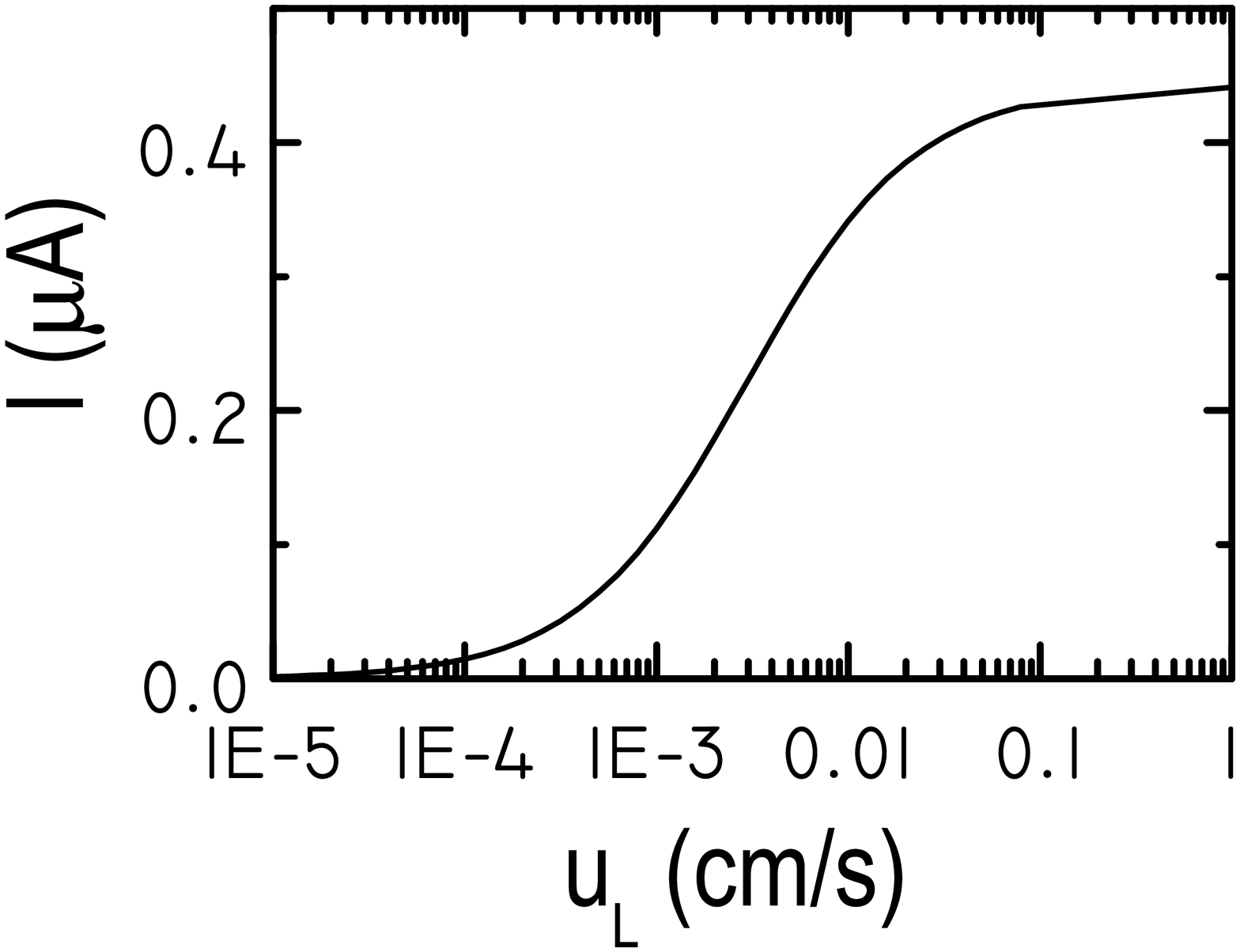}
\caption{Induced velocity with flow speed for parameter values as in Fig. \ref{repulwbpt5kappaapt2} on a log scale for comparison with the behavior predicted by \cite{shankar2} as shown in Fig. \ref{shankarfig}.}
\label{comparewithanalytics}
\caption{Current $I$ as a function of flow speed $u_L$. Figure provided by \cite{shankar2}.}
\label{shankarfig}
\end{figure}

We observe that our simulations (Fig. \ref{comparewithanalytics}) agree well with the theoretical predictions of a saturating dependence of the induced velocity of the carriers on the imposed flow speeds (Fig. \ref{shankarfig}). We further find the current {\em decreases} when $\dot \gamma$ passes a threshold value $\dot \gamma_c$, and that $\dot \gamma_c$ is {\em smaller} for smaller $W_B$. This observation demands a theoretical explanation which at present we do not have. We hope this will motivate further experiments to check this prediction.
It is significant that we are able to observe this type of ``Brownian drag'' and transfer of momentum even in the
absence of momentum conservation and in the extreme limit of no inertia.
Also, though strictly speaking fluctuation dissipation relations are valid only in equilibrium, a naive extension of FDT as in \cite{shankar2} to the flowing case would imply a decrease in the rate of relaxation with increasing flow rate, and hence a ``loss of grip'' and a velocity weakening, which is what we find.

\acknowledgements

We thank S. Ghosh for useful discussions and providing
figure \ref{shankarfig}. MD acknowledges CSIR, India for support, and SR and GA thank
the DST, India for support through the Centre for Condensed Matter Theory.
AKS thanks DST for financial support of his research on carbon nanotubes.

\end{document}